\documentstyle[preprint,aps,epsfig]{revtex}
\draft
\begin{document}
\preprint{Phys. Rev. C in press.}
\title{Isospin-Dependence of $\pi^-/\pi^+$ Ratio and Density-Dependence of Nuclear Symmetry Energy}
\bigskip
\author{\bf Bao-An Li\footnote{email: Bali@astate.edu}}
\address{Department of Chemistry and Physics\\
P.O. Box 419, Arkansas State University\\
State University, Arkansas 72467-0419, USA}
\maketitle

\begin{quote}
The $\pi^-/\pi^+$ ratio is examined as a function of isospin asymmetry and 
beam energy for central collisions induced by neutron-rich nuclei within an isospin-dependent 
hadronic transport model. The $\pi^-/\pi^+$ ratio is found to increase with the isospin asymmetry, 
but decrease with the beam energy of the reaction. For neutron-rich systems, such as,
$^{124}Sn+^{124}Sn$ and $^{132}Sn+^{124}Sn$, the sensitivity of $\pi^-/\pi^+$ ratio
to nuclear symmetry energy is almost beam energy independent.  
\end{quote}

\newpage
Nuclear reactions using rare isotopes has opened up several new frontiers in nuclear 
sciences\cite{tanihata,liudo,lkb98}. 
In particular, heavy rare isotopes currently 
available at several radioactive beam facilities in the world 
and the more energetic ones to be available at the planned Rare Isotope 
Accelerator (RIA) provide a unique opportunity to 
explore novel properties of dense neutron-rich matter that was not in reach 
in terrestrial laboratories before. This exploration will reveal crucial
information about the equation of state (${\rm EOS}$) of 
neutron-rich matter\cite{sjy,xu,betty,li00,ditoro}. 
Within the parabolic approximation (see e.g., \cite{wiringa88}), 
the ${\rm EOS}$ of neutron-rich matter of isospin asymmetry 
$\delta\equiv (\rho_n-\rho_p)/(\rho_n+\rho_p)$ can be written as
\begin{equation}\label{ieos}
e(\rho,\delta)= e(\rho,0)+E_{sym}(\rho)\delta^2+{\cal O}(\delta^4)
\end{equation}
where $e(\rho,0)$ is the energy per nucleon in isospin symmetric 
nuclear matter. The density-dependent nuclear symmetry energy 
$E_{sym}(\rho)$, especially at high densities, is very uncertain and 
has been a subject of extensive research with various 
microscopic and/or phenomenological models over the last few decades.
Besides being important for studying the structure of radioactive nuclei near neutron and proton driplines, 
the density dependence of nuclear symmetry energy  
has many profound consequences for several key issues in 
astrophysics\cite{kut,lat00,pra97,bom01,hor}. It is therefore an important theoretical 
challenge to make reliable predictions on which experimental observables can be used as sensitive 
probes of the $E_{sym}(\rho)$, thus constraining the isospin dependence of the nuclear EOS.
Within a hadronic transport model it was recently shown that the $\pi^-/\pi^+$ ratio is a particularly useful 
probe for the high density behavior of the $E_{sym}(\rho)$\cite{li02}. 
For central collisions of $^{132}Sn+^{124}Sn$ a significant sensitivity of the $\pi^-/\pi^+$ to the variation of $E_{sym}(\rho)$
was found. How does this sensitivity depend on the isospin asymmetry of the reaction system?
In this Brief Report, we investigate this question. An answer to it is not only 
interesting in its own right but also useful for planning future experiments\cite{bill}.

It is well known that the $\pi^-/\pi^+$ ratio in heavy-ion collisions depends strongly on the isospin asymmetry of the
reaction system, see, e.g., \cite{ben79,nag81,har85,sto86}. It is also qualitatively easy to understand why this dependence
can be used to extract crucial information about the {\rm EOS} of neutron-rich matter. On one hand, within the $\Delta$
resonance model for pion production from first-chance independent nucleon-nucleon collisions\cite{sto86}, the primordial $\pi^-/\pi^+$ ratio 
is $(5N^2+NZ)/(5Z^2+NZ)\approx (N/Z)^2$. It is thus a direct measure of the isospin 
asymmetry $(N/Z)_{dense}$ of the dense matter in the participant region of heavy-ion collisions. It was shown that the $(N/Z)_{dense}$ 
is uniquely determined by the high density behaviour of the nuclear symmetry energy\cite{li02}. Therefore, the $\pi^-/\pi^+$ ratio
can be used to probe sensitively the {\rm EOS} of neutron-rich matter. 
On the other hand, within the statistical model for pion production\cite{nature}, the $\pi^-/\pi^+$ ratio is proportional to 
${\rm exp}\left[(\mu_n-\mu_p)/T\right]$, 
where T is the temperature, $\mu_n$ and $\mu_p$ are the chemical potentials of neutrons and protons, respectively. 
At modestly high temperatures ($T\geq 4$ MeV), the difference in the neutron and proton chemical potentials can be written as\cite{thermal}
\begin{equation}
\mu_n-\mu_p=V^n_{asy}-V^p_{asy}-V_{Coulomb}+T\left[{\rm ln}\frac{\rho_n}{\rho_p}+\sum_m\frac{m+1}{m}b_m(\frac{\lambda_T^3}{2})^m(\rho^m_n-\rho^m_p)\right],
\end{equation}
where $V_{Coulomb}$ is the Coulomb potential for protons, $\lambda_T$ is the thermal wavelength of a nucleon and $b'_m$s are the inversion 
coefficients of the Fermi distribution function\cite{thermal}. The difference in neutron and proton symmetry 
potentials $V^n_{asy}-V^p_{asy}=2v_{asy}(\rho)\delta$, where the function 
$v_{asy}(\rho)$ is completely determined by the density-dependence of the symmetry energy\cite{li02}. It is seen that the kinetic 
part of the difference $\mu_n-\mu_p$ relates directly to the isospin asymmetry $\rho_n/\rho_p$ or $\rho_n-\rho_p$.
Thus within the statistical model too, the $\pi^-/\pi^+$ ratio is sensitive to the $(N/Z)_{dense}$. Moreover, the value of $\pi^-/\pi^+$ ratio 
is affected by the competition of the symmetry and Coulomb potentials which all depend on the isospin asymmetry of the reaction system.
These expectations based on two idealized models illustrate qualitatively the usefulness of the 
$\pi^-/\pi^+$ ratio for investigating the {\rm EOS} of neutron-rich matter. For more quantitative studies, however, advanced transport model
calculations are necessary. 

To investigate quantitatively effects of the isospin asymmetry of the reaction system on the $\pi^-/\pi^+$ ratio, we extend the study 
of ref.\cite{li02} based on a hadronic transport model\cite{ibuu1}. We compare three reaction systems 
$^{112}Sn+^{112}Sn$, $^{124}Sn+^{124}Sn$ and $^{132}Sn+^{124}Sn$ 
in the beam energy range of 200 to 2000 MeV/nucleon. Since we are interested in situations where the average N/Z ratio of the participant 
region is the same as that of the initial reaction system, we concentrate on central collisions at an impact parameter of 1 fm.
To explore the isospin-dependence of the nuclear {\rm EOS}, two phenomenological equations of states with the same symmetric part $e(\rho,0)$ 
corresponding to a compressibility $K_{0}=210$ MeV but different forms for the symmetry energy $E_{sym}(\rho)$ are used.
Above the normal nuclear matter density $\rho_0$, the density dependence of $E_{sym}(\rho)$ is very controversial. 
Theoretical results with various models and interactions can be classified into two approximately equally large groups, 
i.e., a group where the $E_{sym}(\rho)$ rises monotonously and one in which it begins to fall above about $2\rho_0$, 
see e.g., \cite{bom01,stone,mar}. 
As in ref.\cite{li02}, we use the following two representative parameterizations 
\begin{equation}\label{esym}
E^a_{sym}(\rho)\equiv E_{sym}(\rho_0)u
\end{equation}
and
\begin{equation}
E^b_{sym}(\rho)\equiv E_{sym}(\rho_0)u\cdot\frac{3-u}{2},
\end{equation}
where $u\equiv\rho/\rho_0$ is the reduced density.  
The linearly increasing $E^a_{sym}(\rho)$ is the typical prediction of the Relativistic Mean Field (RMF) theory\cite{hor,serot}. 
The $E^b_{sym}(\rho)$ first increases with density, then starts to decrease above $1.5\rho_0$ and becomes zero at $3\rho_0$. 
It approximates the prediction of Hartree-Fock calculations using several Skyrme and/or Gogny effective interactions\cite{bom01,stone,mar}. 

Shown in Fig.\ 1 are typical results for the three reactions.
In the left panels we examine the average isospin asymmetry $(n/p)_{\rho\geq \rho_0}$ of the whole space where the local densities 
are higher than the $\rho_0$. Effects of the symmetry energy is clearly seen after about 10 fm/c when a sufficiently high compression 
has been reached. The $E^a_{sym}(\rho)$ 
($E^b_{sym}(\rho)$) makes the high density region more neutron poor (rich) compared to the initial system due to the continuous isospin
fractionation\cite{li02}. This effect is stronger for the more neutron-rich systems as one expects.    
Shown in the right panels are the $(\pi^-/\pi^+)_{like}$ 
ratio
\begin{equation}
(\pi^-/\pi^+)_{like}\equiv \frac{\pi^-+\Delta^-+\frac{1}{3}\Delta^0}
{\pi^++\Delta^{++}+\frac{1}{3}\Delta^+}.
\end{equation} 
This ratio naturally becomes the final $\pi^-/\pi^+$ ratio at the freeze-out after all $\Delta$ resonances have decayed. 
In the early stage of the reaction, the $(\pi^-/\pi^+)_{like}$ ratio is rather high for the neutron richer systems 
because of the large numbers of neutron-neutron scatterings near the surfaces where the neutron skins of the colliding nuclei overlap. 
It saturates after about 25 fm/c for all three systems. The higher $\pi^-/\pi^+$ ratio with the symmetry energy $E^b_{sym}(\rho)$
reflects directly the higher $n/p$ ratio reached in the high density region. 
To further examine the system dependence, we plot the 
saturated final $\pi^-/\pi^+$ ratio as a function $(N/Z)_{system}$ in the left window of Fig.\ 2. As references, we also plotted the
$(N/Z)$ and $(N/Z)^2$. First of all, it is seen that the $\pi^-/\pi^+$ ratio falls far below the first-chance $\Delta$ resonance model
prediction $(N/Z)^2$. This is because of the pion reabsorptions and rescatterings ($\pi+N\leftrightarrow \Delta$ and 
$N+\Delta\leftrightarrow N+N$) which reduce the sensitivity of the $\pi^-/\pi^+$ ratio to the $(N/Z)_{system}$.
Moreover, what is more important for the $\pi^-/\pi^+$ ratio is the local, changing $n/p$ ratio during particularly the compression phase of the reaction.
Secondly, the effect of the symmetry energy on the $\pi^-/\pi^+$ ratio is seen to increase slightly as one goes from $^{112}Sn+^{112}Sn$ 
to $^{124}Sn+^{124}Sn$. As one goes further to the $^{132}Sn+^{124}Sn$ system, the effect remains at about 15\%. Thus, as far as the
$\pi^-/\pi^+$ probe of the symmetry energy is concerned, neutron-rich stable beams, such as $^{124}Sn$, seem to be sufficient. It is thus 
tempting to suggest a re-analysis of existing $\pi^-/\pi^+$ data from heavy-ion collisions at GSI energies.   
The beam energy dependence of the  $\pi^-/\pi^+$ ratio for the reaction of $^{132}Sn+^{124}Sn$
is shown in the right window of Fig.\ 2. While the $\pi^-/\pi^+$ ratio decreases with the
increasing beam energy, its sensitivity to the symmetry energy remains about the same. Similar results are found also for the other two reaction systems. 
The decreasing $\pi^-/\pi^+$ ratio is mainly because of the increasingly important contributions of pions from second-chance 
nucleon-nucleon collisions as the beam energy increases. If a first chance nucleon-nucleon collision converts a neutron to a proton by producing a $\pi^-$,
subsequent collisions of the still energetic proton can convert itself back to a neutron by producing a $\pi^+$. Eventually, at very high energies
the sequential multiple nucleon-nucleon collisions will lead to a $\pi^-/\pi^+\approx 1$.  

In summary, within an isospin-dependent hadronic transport model using two representative density-dependent symmetry energy 
functions predicted by many-body theories, it is shown that the $\pi^-/\pi^+$ ratio increases with the isospin asymmetry, 
but decrease with the beam energy of the reaction. For neutron-rich systems, such as,
$^{124}Sn+^{124}Sn$ and $^{132}Sn+^{124}Sn$, the sensitivity of $\pi^-/\pi^+$ ratio
to nuclear symmetry energy is almost beam energy independent.  

The author would like to thank W.G. Lynch for suggesting me to conduct this research and useful discussions.
This work was supported in part by the National Science Foundation Grant 
No. PHY-0088934 and Arkansas Science and Technology Authority Grant No. 00-B-14.

\newpage

\begin{figure}[htp] 
\vspace{3.5cm}
\centering \epsfig{file=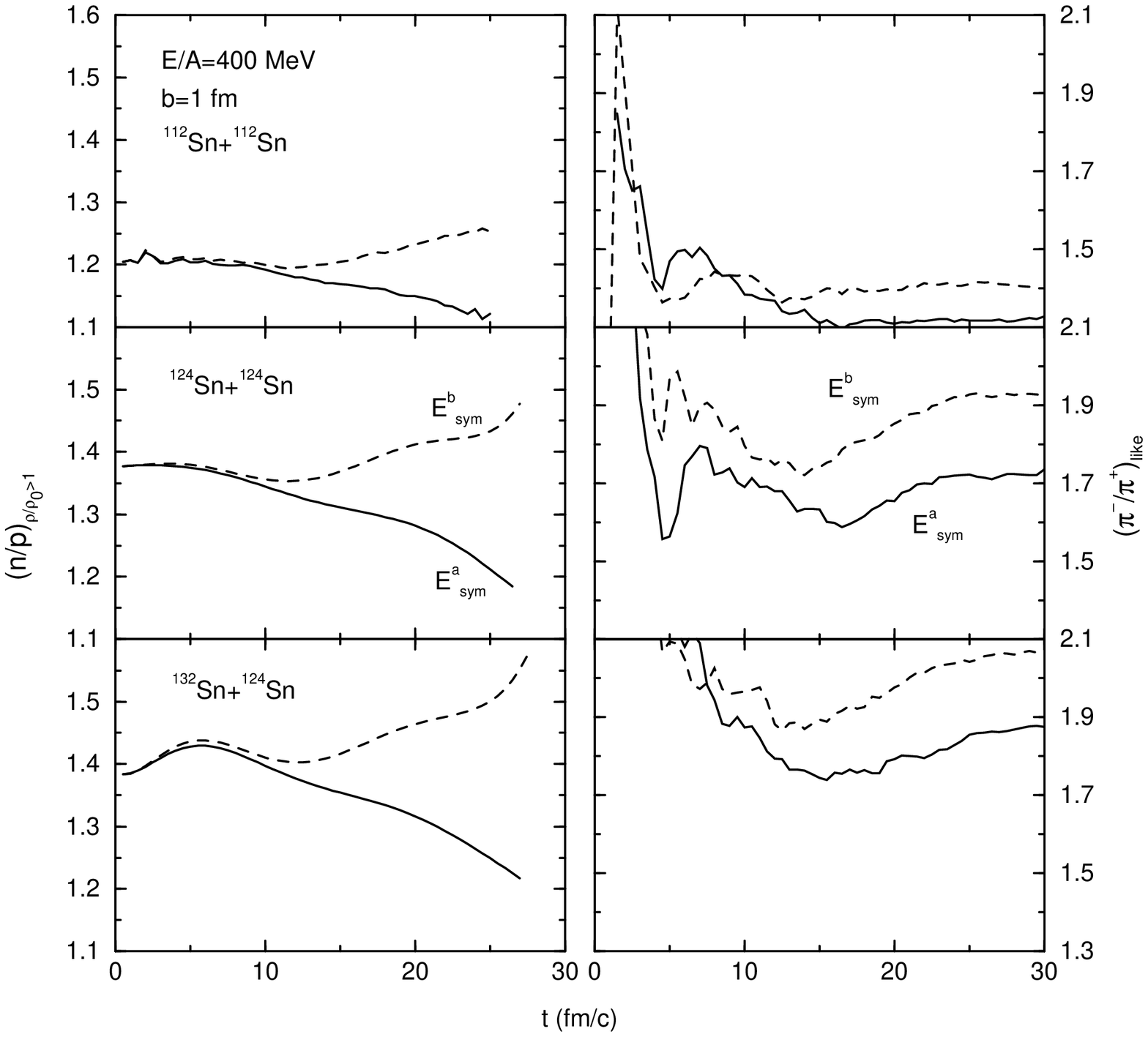,width=12cm,height=13cm,angle=-90} 
\vspace{1.cm}
\caption{Left panels: average neutron/proton ratio in the whole space with densities higher than the normal nuclear matter density as a function of time.
Right panels: the $(pi^-/\pi^+)_{like}$ ratio as a function of time for the three reactions. The solid (dashed) lines are the results using the symmetry
energy $E^a_{sym}(\rho)$ ($E^b_{sym}(\rho)$)}. 
\label{fig1}
\end{figure}

\begin{figure}[htp] 
\vspace{3.5cm}
\centering \epsfig{file=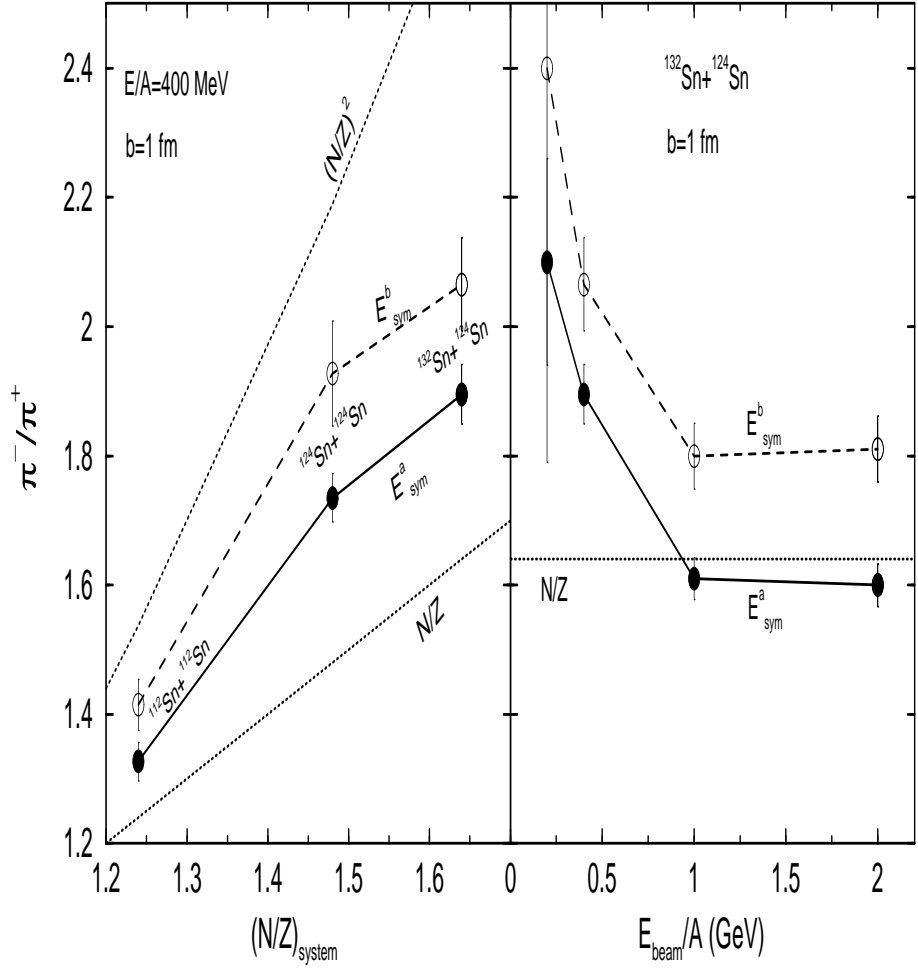,width=12cm,height=13cm,angle=-90} 
\vspace{1.cm}
\caption{The $(pi^-/\pi^+)$ ratio as a function of the isospin asymmetry (left window) and beam energy (right window) of the reaction system} 
\label{fig2}
\end{figure}

\end{document}